\documentclass[12pt,preprint]{aastex}

\usepackage{natbib}

\shorttitle{Occultation by 87 Sylvia }

\shortauthors{Lin et al.}

\begin{document}

\title{A Close Binary Star Resolved from Occultation by 87 Sylvia
      }

\author{
Chi-Long Lin\altaffilmark{1}, 
Zhi-Wei Zhang\altaffilmark{2}, 
W. P. Chen\altaffilmark{2},
Sun-Kun King\altaffilmark{3}, 
Hung-Chin Lin\altaffilmark{2}, 
F.~B.~Bianco\altaffilmark{4,5},
M.~J.~Lehner\altaffilmark{3,5},
N.~K.~Coehlo\altaffilmark{6},
J.-H.~Wang\altaffilmark{3,2},
S.~Mondal\altaffilmark{2},
C.~Alcock\altaffilmark{5},
T.~Axelrod\altaffilmark{7},
Y.-I.~Byun\altaffilmark{8},
K.~H.~Cook\altaffilmark{9},
R.~Dave\altaffilmark{10},
I.~de~Pater\altaffilmark{11},
R.~Porrata\altaffilmark{11},
D.-W.~Kim\altaffilmark{8},
T.~Lee\altaffilmark{3},
J.~J.~Lissauer\altaffilmark{12},
S.~L.~Marshall\altaffilmark{13,9},
J.~A.~Rice\altaffilmark{6},
M.~E.~Schwamb\altaffilmark{14},
S.-Y.~Wang\altaffilmark{3}, and
C.-Y.~Wen\altaffilmark{3}
      }
\altaffiltext{1}{Exhibit Division, National Museum of Natural Science, 1 Kuan-Chien Rd., Taichung 404, Taiwan}
\altaffiltext{2}{Institute of Astronomy, National Central University, 300 Jhongda Rd, Jhongli,  32054 Taiwan} 
\altaffiltext{3}{Institute of Astronomy and Astrophysics, Academia Sinica, P.O. Box 23-141, Taipei 106, Taiwan}
\altaffiltext{4}{Department of Physics and Astronomy, Univ. of Pennsylvania, 209 South 33rd Street, Philadelphia, PA 19104}
\altaffiltext{5}{Harvard-Smithsonian Center for Astrophysics, 60 Garden Street, Cambridge, MA 02138}
\altaffiltext{6}{Department of Statistics, University of California Berkeley, 367 Evans Hall, Berkeley, CA 94720}
\altaffiltext{7}{Steward Observatory, 933 North Cherry Avenue, Room N204 Tucson AZ 85721}
\altaffiltext{8}{Department of Astronomy, Yonsei University, 134 Shinchon, Seoul 120-749, Korea}
\altaffiltext{9}{Institute of Geophysics and Planetary Physics, Lawrence Livermore National Laboratory, Livermore, CA 94550}
\altaffiltext{10}{Initiative in Innovative Computing, Harvard University, 60 Oxford St, Cambridge, MA 02138}
\altaffiltext{11}{Department of Astronomy, University of California Berkeley, 601 Campbell Hall, Berkeley CA 94720}
\altaffiltext{12}{Space Science and Astrobiology Division 245-3, NASA Ames Research Center, Moffett Field, CA, 94035}
\altaffiltext{13}{Kavli Inst. for Particle Astrophysics and Cosmology, 2575 Sand Hill Road, MS 29, Menlo Park, CA 94025}
\altaffiltext{14}{Division of Geological and Planetary Sciences, California Institute of Technology, 1201 E. California Blvd., 
           Pasadena, CA 91125}

\begin{abstract}
The star BD+29\,1748 was resolved to be a close binary from its occultation by the asteroid 87 
Sylvia on 2006 December 18 UT.  Four telescopes were used to observe this event at two sites separated 
by some 80~km apart.  Two flux drops were observed at one site, whereas only one flux drop 
was detected at the other.  From the long-term variation of Sylvia, we inferred the probable shape 
of the shadow during the occultation, and this in turn constrains the binary parameters:~the two 
components of BD+29\,1748 have a projected separation of 0\farcs097--0\farcs110 on the sky 
with a position angle 104\degr--107\degr.  The asteroid was clearly resolved with a size scale 
ranging from 130 to 290~km, as projected onto the occultation direction.  No occultation was detected 
for either of the two known moonlets of 87 Sylvia.  
\end{abstract}

\keywords{asteroids; binaries: general; techniques: high angular resolution; occultations}

\section{INTRODUCTION}

Stellar occultation provides a way to get high angular-resolution information of a celestial object.
When the occulting object is well known (e.g., the lunar limb), an occultation event, 
when observed with fast photometry \citep{war88}, can be used to 
study the background object, e.g., to resolve a close binary \citep{tho06} or to measure the stellar 
diameter \citep{rid79}.  If the object being occulted is reasonably understood (e.g., a point star),
the properties of the foreground object can be inferred, e.g., the planetary atmosphere, rings, or  
the size and shape of an asteroid \citep{eli79}.
The occultation technique also found applications in geodesy if the astrometry of both objects in 
an occultation event is well secured \cite[e.g.,][]{hen58}.  In particular, a stellar occultation by an 
asteroid, if recorded by a group of geographically distributed observers, can yield not only the 
size but also the overall shape of the asteroid.  For instance, the asteroid 216 Kleopatra was 
depicted as a long scraggly bar \citep{dun91} before its dog-bone shape was revealed with radar 
observations \citep{ost00}.  Furthermore, a 3-D model can be achieved by combining data collected 
from several independent occultation events of the same asteroid \citep{mak05}.

The asteroid 87 Sylvia is a large, X-type, outer main-belt asteroid, with dimensions of $385 \times 265 \times 230 
\pm 10$~km \citep{mar05}.  Its binarity was suspected early on by its light variations \citep{pro94,kaa02}, 
before direct imaging observations by the Keck II telescope revealed a companion \citep{bro01}.
Another moon was later found \citep{mar05}, making Sylvia the first asteroid known to have two moonlets.
Direct mass and density determinations are hence possible.  Its density of $1.2 \pm 0.1$~gm~cm$^{-3}$ 
suggests a porous, ``rubble-pile'' internal structure \citep{mar05}.  The larger moon, Romulus, has a diameter of 
$\sim18$~km, with an orbital distance of $1365\pm5$~km from Sylvia, whereas the other moon Remus 
measures 7~km across and has an orbital distance of $706\pm5$~km.  It is suggested that the system was 
formed through a recollection of fragments from a disruptive collision.

Here we report the observation of a stellar occultation event of BD$+$29\,1748 (SAO\,80166, 
RA=08:25:01.66, DEC= +28:33:55.3, J2000) by 87 Sylvia on 18 December 2006.  
The asteroid was clearly resolved while no occultation was detected due to either 
of the 2 moonlets.  However, the background star BD$+$29\,1748 was found to be a close binary.
In our private communication with Mr. Dave Herald, we learned that up to the end of 2006, there were 24, 
including ours, binary discoveries among about 1000 successful observations of asteroid occultation events, 
for which the separation and position angle of the binary components, were successful determined.  Our 
observations, however, present an interesting case where the binary was resolved only at one site.  
Determination of binary parameters would have been impossible, but here we included the tri-axial 
dimension \citep{mar05,joh05} and long-term brightness variation \citep{ham07,beh07} of Sylvia into 
consideration to constrain the size and shape of the shadow, thereby allowing an estimate of the angular 
separation and position angle of the binary.  
\S~2 describes the observations by 4 telescopes at two sites in central Taiwan.
\S~3 presents the analysis of the light curves and derivation of the binary separation and orientation 
of the two stellar components in BD+29\,1748.  \S 4 outlines the conclusion of our study.

\section{OBSERVATIONS AND DATA ANALYSIS }

The shadow of the asteroid 87 Sylvia projected by BD+29\,1748 was predicted to pass through central Taiwan 
on 2006 December 18 \citep{sat06,pre06}, with a ground speed of 12.14~km~s$^{-1}$.  Four telescopes, three 
at Lulin Observatory and one in Taichung, joined to observe this event.  Two of the telescopes used at 
Lulin are 50~cm/F1.9 TAOS telescopes \citep{leh08}, each equipped with an SI-800 CCD camera, rendering 
a field of view of 3 square degrees.  The TAOS telescopes operate in a shutterless shift-and-pause 
charge transfer mode to achieve a sampling rate of 5~Hz, and are designed to monitor photometry of several hundred 
stars simultaneously for chance occultation by Kuiper belt objects \citep{zha08}.  The TAOS system interrupted its 
routine operation to observe the predicted stellar occultation by 87 Sylvia, in an effort to detect the event 
by multiple telescopes and from different sites.  Another telescope used at Lulin was a 40-cm/F10 
telescope, equipped with an Andor U-42 CCD camera.  This camera observed the Sylvia event with regular imaging, i.e., 
with an exposure between the open and close of a shutter, with a sampling rate of 
approximately 2 seconds.  In Taichung, some 80~km to the northwest of Lulin, an amateur 25~cm/F4 Schmidt-Newton 
telescope equipped with a Watec-902H video camera (33~ms sampling rate), with no spectral filter, was used 
on the rooftop of a resident building.  The video images were digitized and analyzed to obtain the light 
curve of the occultation. 

For the Sylvia event, the built-in clocks of all three telescopes at Lulin Observatory were well calibrated so 
the timing was recorded directly without further adjustment.  For the video data taken in Taichung, however,  
we took frames of the on-screen clock of a calibrated computer before the event and counted the difference 
between the event frame and the reference frame to determine the time of the event.

\begin{figure}
\center
\includegraphics[width=14cm]{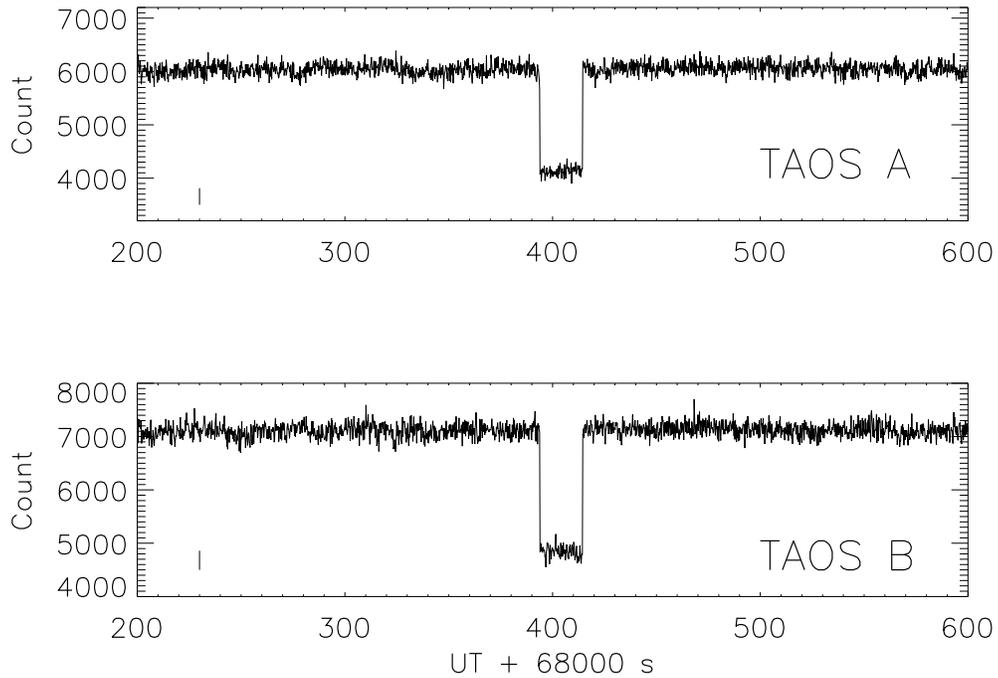}
\caption{The light curves taken by the two TAOS telescopes, each with a sampling time of 0.2~s.  The 
      timing (in seconds) is referenced to the beginning of the day.  The vertical bar in the lower left 
      shows the 3-sigma fluctuation of the data away from the occultation event.  
      }
\label{fig:taos}
\end{figure}

\begin{figure}
\center
\includegraphics[width=8cm]{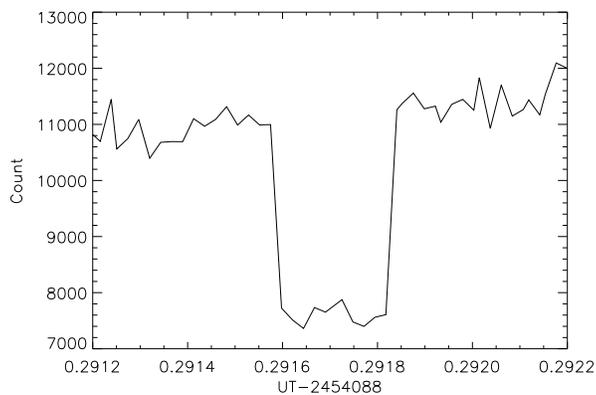}
\caption{Light curve taken by the 40-cm telescope at Lulin, located adjacent to the TAOS telescopes.
          }
\label{fig:slt}
\end{figure}

\begin{figure}
\center
\includegraphics[width=14cm]{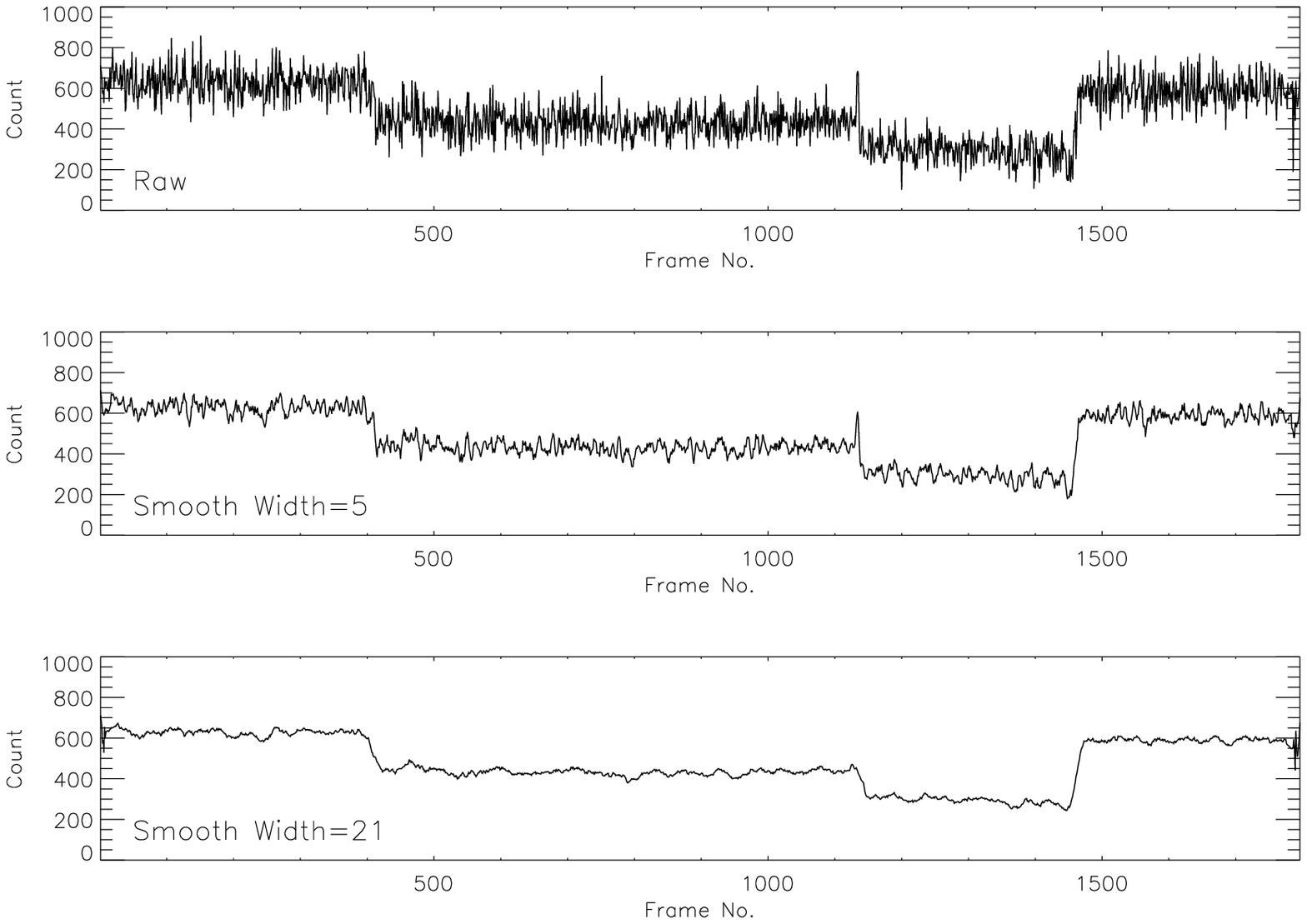}
\caption{({\it Top}) The light curve taken by the video camera in Taichung, with a sampling rate of 
          33 frames per second;  
	  ({\it Middle}) The same light curve smoothed by a running box-car average (equal weighting) 
	    with a width of 5 frames; 
	  ({\it Bottom}) The same but smoothed with a width of 21 frames. 
      } 
\label{fig:taichung}
\end{figure}

Fig.~\ref{fig:taos} shows the light curves taken by the TAOS telescopes.  Both TAOS telescopes detected a 
$\Delta m = 0.42 \pm 0.07$~mag flux drop lasting for 20.4~s, which corresponds to a size scale of 247.66~km. 
The 40-cm telescope, located some 20~m away from the TAOS telescopes, detected a $\Delta m = 0.39$~mag flux 
drop lasting for 20~s (Fig.~\ref{fig:slt}), consistent with the TAOS measurements.  All three telescopes at 
Lulin therefore must have seen the same occultation event.  The slight difference 
between the TAOS and 40-cm results may be attributed to different filters used; the 40-cm telescope used 
a standard V-band filter, whereas each TAOS system used a very broad-band (500--700~nm) filter.   
The flux drops, however, are inconsistent with the predicted value of $\Delta m = 2.7$~mag \citep{pre06,sat06}.
The case is vindicated in the Taichung data, shown in Fig.~\ref{fig:taichung}, 
for which, in addition to a $\Delta m = 0.42$~mag drop lasting for 23.9~s, corresponding to 290.15~km, 
there was a brief reappearance, followed by a second, more appreciable flux drop of $\Delta m = 0.93$~mag 
lasting for 10.83~s (131.45~km).  The instruments used, and the observational results are summarized 
in Table~\ref{tab:1} and \ref{tab:2}. 

The asteroid 87 Sylvia was clearly resolved.  The duration of occultation multiplied by the shadow 
speed gives the size of the asteroid projected onto the occultation direction.  Three such chords, 
ranging from 130 to 290~km, were measured in our case, all within the known triaxial dimensions 
of the asteroid.  No occultation was detected for either of the two moonlets.  Moreover, it is obvious 
that the star under occultation, BD+29\,1748, is a binary. 

\begin{deluxetable}{cccccc}
\tablecaption{Instrument Parameters \label{tab:1} }
\tablewidth{0pt}
\tabletypesize{\footnotesize}
\tablehead{ \colhead{ Site}     &  \colhead{ Coordinates }  & \colhead{ Elevation } &  \colhead{ Telescope } & 
              \colhead{ Detector }  & \colhead{ Data Storage }\\ }
\startdata
Lulin    &  120\degr 52\arcmin 25\arcsec E 23\degr 28\arcmin 07\arcsec N  & 2,862 m   & 0.5 m/F1.9 &
           SI 800     &  Personal Computer \\
         &                                                          &           & 0.4 m/F10  &
           U42        & Personal Computer \\
Taichung & 120\degr 39\arcmin 00\arcsec E 24\degr 06\arcmin 48\arcsec N   & 68 m      & 0.25 m/F4     &
           Watec-902H & Digital Camcorder \\
\enddata
\end{deluxetable}

\begin{deluxetable}{llccc}
\tablecaption{Observational Results \label{tab:2} }
\tablewidth{0pt}
\tabletypesize{\footnotesize}
\tablehead{ \colhead{ Site}     &  \colhead{ Immersion ~  Emersion } &  
              \colhead{  Immersion   Emersion }  & \colhead{ Duration } &
	      \colhead{ Flux Drop} \\ 
	                        &   \colhead{h:m:s ~~~~~~~ h:m:s} &  \colhead{h:m:s ~~~~~~~~ h:m:s} & 
		\colhead{s}  & \colhead{ mag } \\}
\startdata
Lulin/TAOS  & 18:59:54 19:00:14.4     & -   & 20.4 $\pm$ 0.2    & 0.41  \\
Lulin/0.4 m   & 18:59:53 19:00:13       & -   & 20 $\pm$ 2        & 0.39  \\
Taichung    & 18:59:57.73 19:00:21.63 & -   & 23.90 $\pm$ 0.033 & 0.42 \\
            &  - & 19:00:21.86  19:00:32.69 & 10.83 $\pm$ 0.033  & 0.93 \\
\enddata
\end{deluxetable}

The BD$+$29\,1748 system has a spectral type of about F8 as given in the SAO catalogue.  
The spectral type was confirmed by a spectrum taken by the one-meter telescope at Lulin 
a few days after the Sylvia event.  The UCAC2 catalogue lists an apparent magnitude of 
V$\sim 9.9$~mag for BD$+$29\,1748. The flux drops for the two components, 0.93 and 0.41~mag, 
therefore lead to 10.42~mag for the primary 
and 10.94~mag for the secondary.  If extinction is negligible, given an absolute magnitude of 
4.0 for an F8 dwarf, the distance to BD$+$29\,1748 should be $\sim 370$~pc.  The secondary, 
if also a dwarf, then should be a G1 star.  With a size of 200--300~km across, the asteroid 
would have an angular size of $\sim 0\farcs1$.  In comparison, either background star has a 
physical size similar to that of the Sun, and therefore would sustain an angular size of about 
0.01 milliarcsecond at its distance, which is much smaller than the asteroid, and could not be resolved 
in our observations.

\section{DERIVATION OF BINARY PARAMETERS }

Because BD$+$29\,1748 was resolved to be a binary pair only at one site, derivation of the binary parameters, 
namely the angular separation and orientation in the sky of the star components, would have been impossible.  
However, because the size and shape of Sylvia has been well measured, it turns out to be feasible to constrain  
the binary geometry with reasonable accuracy from the tri-axial dimension of the asteroid and its long-term 
brightness variation.

\begin{figure}

\center
\includegraphics[width=10cm]{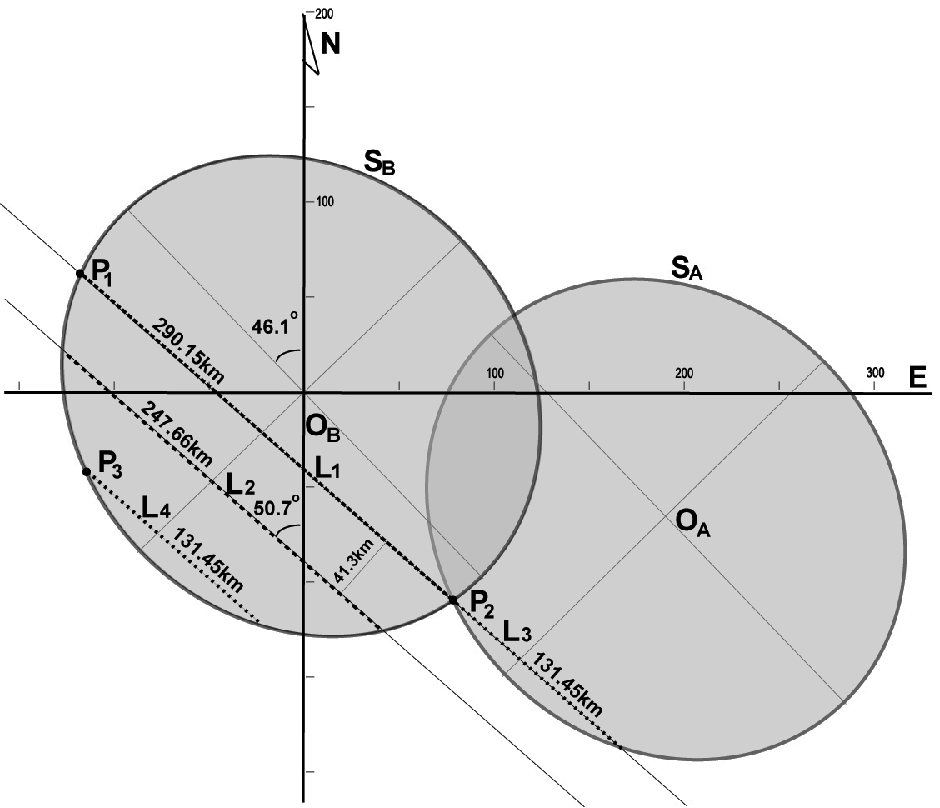}

\vspace{1cm}

\includegraphics[width=10cm]{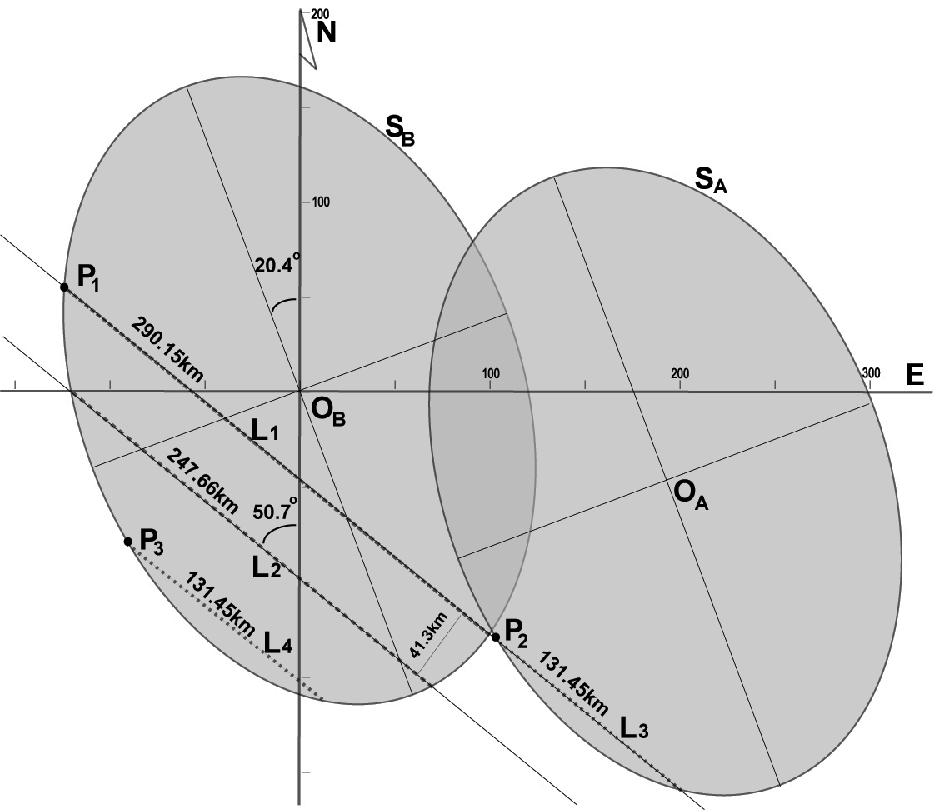}
\caption{The geometry of the asteroid shadow in the two extreme cases.  The orientation of the binary stars 
has a mirrored configuration of the shadow, i.e., with B (secondary) being to the south-east of A (primary). 
}
\label{fig:cases}
\end{figure}

We started out with the shadows of Sylvia illuminated by the primary ($S_A$) and by the companion ($S_B$) 
projected onto the surface of the Earth at the time of occultation (Fig.~\ref{fig:cases})  For simplicity, 
we assume elliptical shadows.  The constraints on the geometry of the shadow ellipses are as follows:

\begin{enumerate} 
 \item The shadow caused by the primary ($S_A$) is identical to that by the secondary ($S_B$).
 \item The shadow was predicted to move with a speed of 12.14~km~s$^{-1}$ in the direction of $309\fdg3$ east of north.     
 \item One site (Taichung) witnessed the occultation events of both the primary and secondary, 
        whereas the other site (Lulin) detected only the event of the secondary.
 \item There are two chords through $S_B$, with lengths of $L_1=290.15$~km (observed in Taichung), and 
    $L_2=247.66$~km (observed at Lulin), and one chord through $S_A$, $L_3=131.45$~km (observed 
    in Taichung) (see Table~2.)  
 \item The distance between $L_1$ and $L_2$ projected onto the occultation direction is 41.3~km.  
\end{enumerate}

Our next step is to estimate the size and shape of the shadow ellipse.  An asteroid varies its brightness 
with its reflecting surface area and phase angle with respect to the Sun as the spinning asteroid orbits 
the Sun.  Given the tri-axial dimensions of $384\times 264 \times 232$~km, Sylvia should 
therefore reach its maximum brightness when its projection becomes a $384\times 264$~km ellipse.  
Sylvia has been measured to have a 0.215984$\pm$0.000002~day period with an amplitude of 
0.273$\pm$0.005~mag \citep{ham07,beh07}.  Tracing back to the time of our observations, we found the 
asteroid's brightness to be 0.26~mag fainter than its maximum.  Thus the cross section of the 
asteroid facing the Earth at the time of observation should be less than 78\% of the $384\times 264$~km ellipse.  
There could be many possible major/minor axes combinations, but the minor axis must be between 232~km 
and 264~km, which in turn dictates the major axis to be 341~km and 300~km, respectively.  
The actual situation must be in between the two extreme cases, a $300 \times 264$~km ellipse 
and a $341\times 232$~km ellipse.  

The next step is to find a line $L_4$ which is parallel to $L_{1}$ and intersects 
$S_B$ with a chord length of 131.45~km.  This chord on $S_{B}$ is to be connected to the one on 
$S_A$ which represents the second occultation observed at Taichung.
The upper-left end $P_3$ of such a chord on $S_{B}$ corresponds to point $P_2$ on $S_{A}$.
The separation and position angle between $S_{A}$ and $S_{B}$, i.e., the binary parameters, can hence be 
readily derived.  The results are demonstrated in Fig.~\ref{fig:cases} for case (1) and case (2), 
and are summarized below.

\noindent
(1) The $300\times 264$~km case  

In the extreme case when the minor axis is 264~km, the length of the major axis should be 300~km.
Note $L_1$ and $L_3$ were both observed in Taichung, but because both $S_A$ and $S_B$ are identical, 
one can find a corresponding chord on $S_B$.  In this configuration, the angle between $L_{1}$ and north 
is 50\fdg7 and the angle between $L_{1}$ and the major axis of $S_B$ is 4\fdg6, so the major axis 
of $S_B$ should be 46\fdg1 counterclockwise from due north as demonstrated in Fig.~\ref{fig:cases}a.
The separation between $S_A$ and $S_B$ is then 227~km or, given the distance of the asteroid of 
2.83016~AU at the time, 0\farcs110 on the sky, with the position angle of the secondary relative to the primary  
to be 107\degr.

\noindent
(2) The $341\times 232$~km case  
 
When the minor axis is at its minimum, i.e., 232~km, the length of the major axis would be 341~km. 
Following the same calculation as above, we obtained the angle between $L_{1}$ and the major axis of 
$S_B$ to be 30\fdg3.  So the major axis of $S_B$ should be 20\fdg4 counterclockwise from north, as 
shown in Fig.~\ref{fig:cases}b.  The separation between shadows $S_A$ and $S_B$ is estimated to 
be 198~km or 0\farcs97, with the position angle of the secondary relative to the primary to be 
104\degr. 

The ranges of binary parameters are values derived from two extreme cases under the assumption of 
an elliptical shadow for 87 Sylvia.  The actual shape of the shadow should be more complex than a simple 
ellipse, whose dimensions, here estimated by light variation of the asteroid, are largely uncertain.  
The elevation of the asteroid (and hence the star) during occultation was 84\degr, i.e., close to zenith, 
so the shadow distortion due to the curvature of the Earth's surface should 
be neglible.   
Note that for each disappearance and reappearance in the light curve recorded in Taichung, which 
is noisy but with a fast sampling rate, as seen in the smoothed light curves (Fig.~\ref{fig:taichung}), the signal 
changed slower than if the occultation would have occurred perpendicular to the local asteroid limb.   
This suggests that the passage of the shadow ellipse was oriented in such a way that the contact 
angle --- the angle relative to the normal at the contact point of the occultation (zero degrees 
for a ``head-on'' occultation, and close to $\pm 90 \degr$ for a grazing event) --- is substantially larger than 0.
As seen in Fig.~\ref{fig:cases}, this is very likely the case, though the actual contact angle is not 
certain because of the irregular shape/limb of the asteroid.  

\section{CONCLUSION }

Occultation by the asteroid 87 Sylvia revealed that the star BD$+$29\,1748 consists of a close pair.  The primary 
is an F8 star with an apparent magnitude of m$_{\rm V}=10.42$, and the secondary, possibly a G1 star, has m$_{\rm V}=10.94$~mag.   
The projected separation between the primary and the secondary is 0\farcs097--0\farcs110, with a position angle
of 104\degr--107\degr.  The asteroid was clearly resolved, while no occultation was detected for either of the
two known moonlets.  

\acknowledgments  We thank David W. Dunham, Sato Isao, Dave Herald and Mitsuru Soma for helpful discussions.


\begin{thebibliography}{}
%
\bibitem[Behrend(2007)]{beh07}
   Behrend, R. 2007, Curves of rotation of asteroids and comets, \\ 
   {\verb http://obswww.unige.ch/~behrend/page1cou.html\#000054 }
\bibitem[Brown(2001)]{bro01}
   Brown, M. E., \& Margot, J. L. 2001, IAUC~7588
\bibitem[Dunham et al.(1991)]{dun91}
   Dunham, D. W., Osborn, W., Williams, G., Brisbin, J., Gada, A., Hirose, T., Maley, P., Povenmire, H., 
   Stamm, J., Thrush, J., Aikman, C., Fletcher, M., Soma, M., Sichao, W. 1991, Lunar and Planet. Inst. 
   Contrib., 765, 54
\bibitem[Elliot(1979)]{eli79}
   Elliot, J. L. 1979, \araa, 17, 445
\bibitem[Hamanowa \& Hamanowa(2007)]{ham07}
   Hamanowa, Hiromi \& Hamanowa, Hiroko, 2007, Asteroid Lightcurve Data File, \\ 
   {\verb http://www2.ocn.ne.jp/~hamaten/00087sylvia-lc.htm }
\bibitem[Henriksen et al.(1958)]{hen58}
   Henriksen, S. W., Genatt, S. H., Marchant, M. Q., Batchlor, C. D. 1958, \aj, 63, 291
\bibitem[Johnson(2005)]{joh05}
   Johnston, Wm. R. 2005, (87) Sylvia, Romulus, and Remus, \\
   {\verb http://www.johnstonsarchive.net/astro/astmoons/am-00087.html }
\bibitem[Kaasalainen, Torppa, \& Piironen(2002)]{kaa02}
   Kaasalainen, M., Torppa, J., \& Piironen, J. 2002, Icarus, 159, 369 
\bibitem[Lehner et al.(2008)]{leh08}
   Lehner, M. J., Wen, C.-Y., Wang, J.-H., Marshall, S. L., Schwamb, M. E., Zhang, Z.-W., 
   Bianco, F. B., Giammarco, J., Porrata, R., Alcock, C., Axelrod, T., Byun, Y.-I., Chen, 
   W. P., Cook, K. H., Dave, R., King, S.-K., Lee, T., Lin, H.-C., Wang, S.-Y., 2008, arXiv.0802.0303 
\bibitem[Marchis et al.(2005)]{mar05}
   Marchis, F. et al., 2005, \nat, 436, 822
\bibitem[Marchis et al.(2006)]{mar06}
   Marchis, F., Kaasalainen, M., Hom, E. F. Y., Berthier, J., Enriquez, J., Hestroffer, D., Le Mignant, D., \& de Pater,
    I. 2006, ICARUS, 185, 39
\bibitem[Maksym(2005)]{mak05}
   Maksym, P., 2005, in Symposium ESOP XXIV, \\
   {\verb https://www.ursa.fi/esop2005/lecture$\_$10.html }
\bibitem[Mason(2006)]{mas06}
       Mason, B.D., Wycoff, G.L. and Hartkopf, W.I , 2006, The Washington Double Star Catalog, \\ 
   {\verb  http://ad.usno.navy.mil/proj/WDS/ }
\bibitem[Ostro et al.(2000)]{ost00}
    Ostro, S. T. et al., 2000, Science, 288, 836.
\bibitem[Perryman et al.(1997)]{per97}
    Perryman, M. A. C., et al., 1997, \aap, 323, L49.
\bibitem[Preston(2006)]{pre06}
    Preston, S. 2006, Asteroid Occultation Prediction website, \\ 
    {\verb http://www.netstevepr.com/Asteroids/archive/2006/2006_12_si.htm }
\bibitem[Prokof\'{e}va \& Demchik(1994)]{pro94}
    Prokof\'{e}va, V. V., \& Demchik, M. I. 1994, Astron. Lett., 20, 245.
\bibitem[Ridgway et al.(1979)]{rid79}
     Ridgway, S. T., Wells, D. C., Joyce, R. R., \& Allen, R. G. 1979, \aj, 84, 24.
\bibitem[Sato(2006)]{sat06}
    Sato, I. 2006, Asteroid Occultation Prediction website, \\ 
    {\verb http://homepage2.nifty.com/mp6338/occultation/t061218.00087.pdf }
\bibitem[Thompson \& Yeelin(2006)]{tho06}
  Thompson \& Yeelin, 2006, \pasp, 118, 1648 
\bibitem[Warner(1988)]{war88}
   Warner, B. 1988, ``{\it High Speed Astronomical Photometry}'' (Cambridge University Press), Chap.~2
\bibitem[Zhang et al.(2008)]{zha08}
   Zhang, Z. W. et al. 2008, \apjl, 685, L157
%
\end{thebibliography}
\end{document}